\documentclass[%
 reprint,
 amsmath,amssymb]{revtex4-2}

\usepackage{CJK}
\usepackage{graphicx}
\usepackage{dcolumn}
\usepackage{bm}
\usepackage{float}
\DeclareFontShape{OT1}{cmss}{m}{it}{<->ssub*cmss/m/sl}{}

\usepackage{enumerate}

\begin{document}

\begin{CJK*}{}{} 
\title{Hydrogen phase-IV characterization by full account of quantum anharmonicity}
\author{Tommaso Morresi,$^{1\ast}$ 
Rodolphe Vuilleumier,$^{2}$ 
Michele Casula$^{1\ast}$\\~\\}
\affiliation{$^{1}$Institut de Min{\'e}ralogie, de Physique des Mat{\'e}riaux et de Cosmochimie (IMPMC), Sorbonne Universit{\'e}, CNRS UMR 7590,  MNHN, 4 Place Jussieu, 75252 Paris, France\\
$^{2}$PASTEUR, D{\'e}partement de chimie, {\'E}cole normale sup{\'e}rieure, PSL University, Sorbonne Universit{\'e}, CNRS, 75005 Paris, France}%

\begin{abstract}
We devise a 
framework to compute accurate phonons in molecular crystals even in case of strong quantum anharmonicity. 
Our approach is based on the calculation of the static limit of the 
phononic Matsubara Green's function 
from path integral molecular dynamics 
simulations.
Our method enjoys a remarkably low variance, which allows one to compute accurate phonon frequencies after a few picoseconds of nuclear dynamics, and it is further stabilized by the use of appropriate constrained displacement operators.
We applied it to solid hydrogen at high pressure.
For phase III, our predicted infrared (IR) and Raman active vibrons agree very well with experiments.
We then characterize the crystalline symmetry of phase IV by direct comparison with vibrational data and we determine the character and isotopic shift of its Raman and IR vibron peaks.
\end{abstract}

\maketitle
\end{CJK*}

\section{Introduction}
One of the main structural features of hydrogen-rich materials is the presence of large quantum fluctuations affecting 
hydrogen nuclear motion, and leading to strong
anharmonicity. 
Pristine hydrogen is intriguing because of its unexpectedly rich phase diagram
in the high-pressure range~\cite{Mao_1994, Loubeyre_1996, Loubeyre_2002, Eremets_2011,McMinis_2015,Dias_2017,Monacelli_2022}.
However, a precise experimental assessment of 
all its phases is difficult, since hydrogen is a weak X-ray and neutron scatterer.
In fact, the most direct
structural information
comes from Raman and infrared (IR)
techniques, probing 
its vibrational properties.
Hence, the underlying geometries could be deduced by comparing experimental data with 
theoretical predictions.
However, that comparison is hindered 
by nuclear quantum fluctuations, which make harmonic
and perturbative methods unreliable.

In this work, we 
generalize and
extend to molecular solids a framework previously introduced \cite{Morresi_2021} to compute vibrational properties of strongly anharmonic systems.
This goal is achieved by accessing the static limit of the phononic Matsubara Green's function and by imposing a constraint on the quantum displacement operators defining the Green's function.
As application, we determine the vibron modes
of phase III and IV in solid hydrogen, entirely from first principles and with unprecedented accuracy. We then address the open issue of identifying the crystalline symmetry of phase IV.

Previous calculations of the vibrational properties of hydrogen were based on classical molecular dynamics (MD)~\cite{Magdau_2013,Zhang_2018}, which cannot include quantum anharmonicity.
To overcome the MD limitations and include nuclear quantum effects (NQE), more advanced methods have been developed, 
such as the vibrational self-consistent field (VSCF) method~\cite{bowman1978self,monserrat2013anharmonic,azadi2014dissociation} and the stochastic self-consistent harmonic approximation (SSCHA)~\cite{Errea_2013,Bianco_2017,Monacelli_2021}. 
Although they account for NQE,
these approximations heavily rely on the quality of their variational ansatz wavefunction. 
In our framework instead, we fully include
NQE by sampling the exact quantum thermal 
distribution, for a given electronic theory, through path integral MD (PIMD) simulations~\cite{Ceriotti_2010,Mouhat_2017} driven by \textit{ab initio} forces. 
PIMD gives direct access to imaginary-time correlation functions and in particular to the exact phononic Matsubara Green's function~\cite{Mahan_2000},
defined as 
\begin{equation}
G_{ij}(\tau)=- \sqrt{m_i m_j} \langle \mathcal{T} \delta \hat{x}_i (\tau) ~ \delta \hat{x}_j (0) \rangle,
\label{eq:imag_time_matsu_green}
\end{equation}
with $\tau$ the imaginary time, $\mathcal{T}$ the time-ordered operator, $\delta x_{i}$ the displacement of $i$-th Cartesian degree of freedom with respect to its equilibrium position, $m_{i}$ its mass, and the brackets indicate the average over the PIMD distribution. However, while Eq.~(\ref{eq:imag_time_matsu_green}) encloses all information on the vibrational properties of the system, the analytic continuation to extract the phonon spectral function suffers from two main issues: it is numerically ill-defined~\cite{Jarrell_1996} and is plagued by
sampling errors inherent in the finite size of PI trajectories. So far, these drawbacks, together with a usually large computational cost, have severely limited practical PIMD-based phonon calculations in realistic Hamiltonians.

In our framework the above issues are solved by taking the Kubo-transform $\tilde{\mathbf{G}}$ of the Green's function in Eq.~(\ref{eq:imag_time_matsu_green}), obtained by averaging the displacement operators over the whole ring-polymer chain representing a quantum particle in the PIMD framework \cite{Craig_2004}. This corresponds precisely to the static limit of the phononic Matsubara Green's function. 
By calculating the inverse $\tilde{\mathbf{G}}^{-1}$, we show that one can get an accurate estimate of the NQE-renormalized phonon frequencies, without requiring any analytic continuation. We also provide a computational scheme where $\tilde{\mathbf{G}}^{-1}$ can be evaluated with low variance and low bias. This goal is achieved by solving a generalized eigenvalue problem (GEV), where the sampling errors of the displacement autocorrelation function are compensated by those of the velocity autocorrelation. 

Nevertheless, the calculation of vibrational properties of molecular solids is hampered by the simultaneous presence of intramolecular and lattice degrees of freedom.  
\begin{figure*}
\includegraphics[scale=0.81]{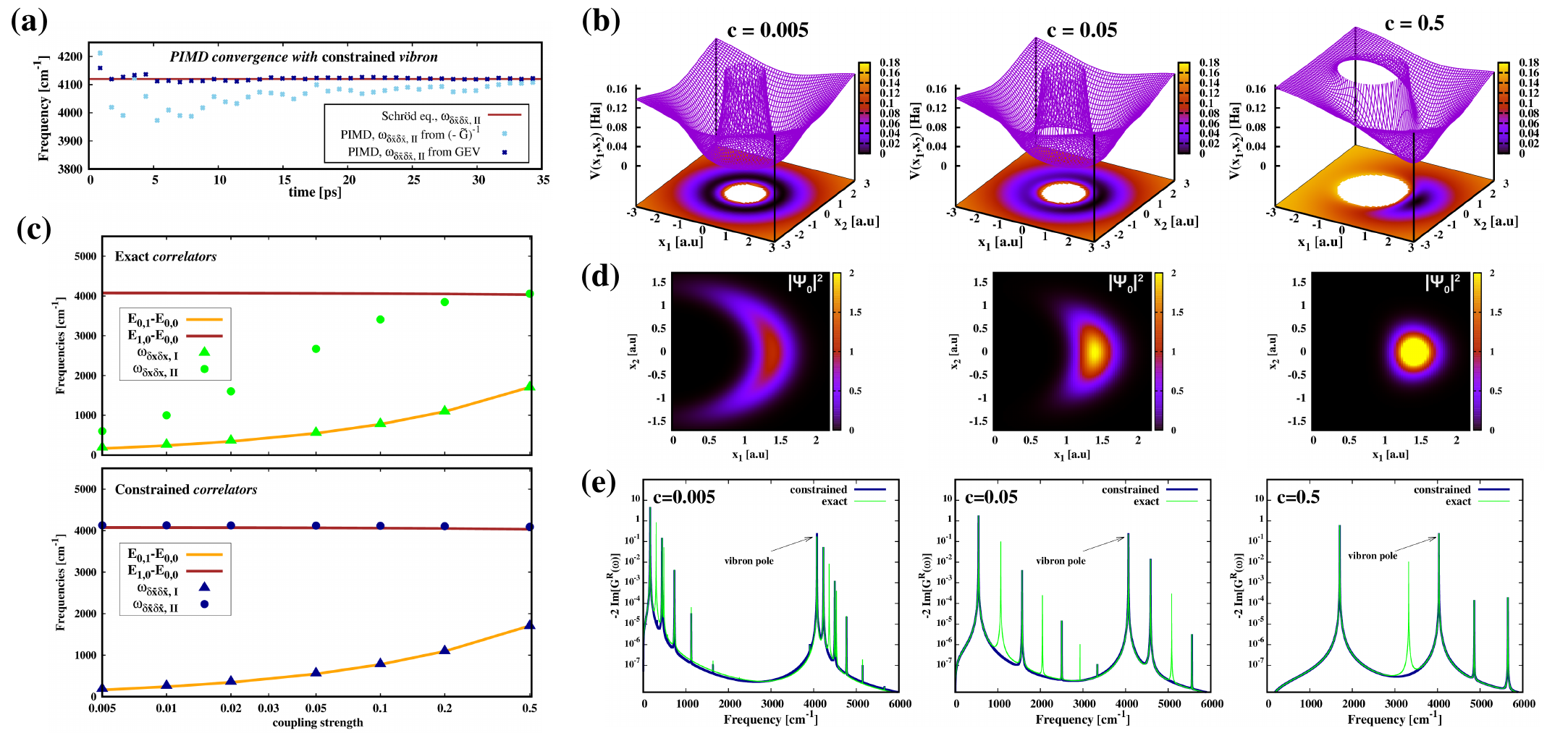}
\caption{2D dimer model. (a) Convergence of the PIMD eigenvalue for the vibron frequency ($\omega_{\delta \bar{x} \delta \bar{x}, \ II}$) of the dimer model in Eq.~(\ref{eq:2dpot_main}) with $c=$ 0.05 as a function of the PIMD evolution.
The GEV frequencies (dark-blue points) converge after a few picoseconds to the solution obtained by solving numerically the corresponding Schr{\"o}dinger equation. (b) Potential shape for three different values of the parameter $c$ in Eq.~(\ref{eq:2dpot_main}): left column c=0.005; central column c=0.05; right column c=0.5. (c) GEV frequencies (Eq.~(\ref{eq:gev})) with exact unconstrained $\delta \hat{\mathbf{x}}$ (top panel) and constrained displacements (lower panel) as a function of the coupling strength $c$ (Eq.~(\ref{eq:2dpot_main})). (d) Square modulus of the ground state wave function ($|\Psi_0|^2$) for the three values of $c$ in panel (b). (e) Imaginary part of the retarded Green's functions for the dimer model of Eq.~(\ref{eq:2dpot_main}) obtained by the numerically exact solution (light green) and with the \textit{constrained} (blue) displacement operators for c=0.005, c=0.05 and c=0.5. The effect of constraints in Eq.~(\ref{eq:dxdx_constrained}) is to suppress some of the rotational poles of the unconstrained Green's function. The small imaginary part in the denominator of the retarded Green's function, corresponding to the width of the peaks, is set equal to 2 cm$^{-1}$.}\label{fig:model_pot}
\end{figure*}
Indeed, while it has been shown~\cite{Morresi_2021} that this approach accurately reproduces the exact spectrum for various model Hamiltonians even in the strong anharmonic limit, its
reliability deteriorates for systems made of rotating molecules, such as the molecular phases of solid hydrogen~\cite{Van_1983}. These systems are characterized by the presence of intramolecular libration and vibron modes. In case of large librations, it turns out that the vibron mode is systematically underestimated, as the $\tilde{\mathbf{G}}^{-1}$ eigenvalues are biased by the hybridization between these modes.
In this paper, we demonstrate how to fix 
this problem by introducing constraints on the displacement operators in Eq.~(\ref{eq:imag_time_matsu_green}), that 
effectively reduce the phonon Fock space to the most relevant sectors. 
After benchmarking the performances of our
constrained PIMD (cPIMD) framework (Sec.~\ref{framework}) against the exact solution 
for a rotor model, which represents a dimer molecule interacting with an external field (Sec.~\ref{model}), we prove the generality of cPIMD by applying it to phases III and IV of solid hydrogen (Secs.~\ref{phase_III} and \ref{phase_IV}, respectively).

\section{Results}
\subsection{Constrained PIMD framework}
\label{framework}
While the evaluation of the Matsubara Green's function (Eq.~\ref{eq:imag_time_matsu_green}) is challenging, PIMD allows one to compute also Kubo-transformed correlation functions~\cite{Craig_2004}, such as $\tilde{G}_{ij}=- \beta \sqrt{m_i m_j} \langle \delta \tilde{x}_i \delta \tilde{x}_j \rangle$,
where $\delta \tilde{x}_i  = \int_0^\beta \!d\tau ~ \delta \hat{x}_i (\tau) / \beta$ is the time-averaged operator, $\beta = 1/(k_B T)$ is the inverse temperature, and
$\tilde{G}_{ij}$ the Fourier transform of Eq.~(\ref{eq:imag_time_matsu_green}) at zero Matsubara frequency. 
In other words, $\tilde{\mathbf{G}} = -\frac{1}{\mathbf{D}^\textrm{harm}+ \boldsymbol{\Pi}(0)}$, where $\mathbf{D}^\textrm{harm}$ represents the harmonic dynamical matrix of ``bare'' phonons, and $\boldsymbol{\Pi}(0)$ is the static limit of the phonon self-energy. Therefore, by computing the inverse $\tilde{\mathbf{G}}^{-1}$, one can have access to squared phonon frequencies renormalized by static correlations and thus avoid the nuisance of analytically continuing Eq.~(\ref{eq:imag_time_matsu_green}) to real frequencies. Moreover, dealing with $\tilde{\mathbf{G}}^{-1}$ cures also the second drawback of the Green's function approach, namely the impact of finite-sampling errors. Indeed, there is an efficient way to drastically reduce its statistical fluctuations. This is achieved by solving the GEV:
\begin{equation}\label{eq:gev}
\left[ \langle \delta \tilde{\mathbf{x}} \delta \tilde{\mathbf{x}}^T \rangle^{-1} \right]_{ij} W_{jk} = \omega^2_k \left[ \langle \tilde{\dot{\mathbf{x}}} \tilde{\dot{\mathbf{x}}}^T \rangle^{-1} \right]_{ij} W_{jk},    
\end{equation}
where $\langle \tilde{\dot{x}}_i \tilde{\dot{x}}_j \rangle$ is the Kubo-transformed velocity-velocity correlation function, $\omega^2_i$ is the squared frequency of the $i$-th phonon mode, and $W_{ji}$ is the eigenvectors matrix. By extending the maximum localization criterion for determining normal modes in classical MD~\cite{Martinez_2006}, the GEV in Eq.~(\ref{eq:gev}) was first proposed in Ref.~\cite{Morresi_2021} within the PIMD framework. The advantage of using Eq.~(\ref{eq:gev}) is illustrated in Fig.~\ref{fig:model_pot}(a), where the GEV eigenvalues converge remarkably faster than the ones obtained by direct inversion of $\tilde{\mathbf{G}}$. The GEV convergence below a given target error requires one-order-of-magnitude shorter PIMD trajectories, thus opening the avenue to realistic systems applications.  

However, issues arise when treating rotational systems using this approach. This is rationalized by rewriting the Matsubara Green's function at zero frequency by means of its Lehmann representation, such that it reads:
\begin{equation}\label{eq:lehmann_green}
\small
     \tilde{G}_{ij} =\beta \frac{\sqrt{m_i m_j}}{ Z} \sum_{l,m} \langle l | \delta \hat{x}_i | m \rangle \langle m | \delta \hat{x}_j | l \rangle \frac{e^{- \beta E_m}-e^{-\beta E_l}}{E_m-E_l},
\end{equation}
where $Z$ is the canonical partition function, and $E_{l}$ is the energy of the $l$-th quantum eigenstate of the system. From Eq.~(\ref{eq:lehmann_green}), it is apparent that the frequencies $\{\omega_i\}_{i=1,\ldots,N}$, resulting from the diagonalization of $\tilde{\mathbf{G}}^{-1}$ for a system with $N$ degrees of freedom, depend not only on the $N$ principal phonon excitations, but on the full spectrum, with higher $m \rightarrow l$ excitations (overtones) weighted by matrix elements given by $\langle l | \delta \hat{x}_i | m \rangle \langle m | \delta \hat{x}_j | l \rangle (e^{- \beta E_m}-e^{-\beta E_l})$. By definition, the principal phonon modes are associated with the poles of the Green's function with the largest spectral weights. When the weights of the main poles are dominant, this framework is nearly exact. However, for strongly anharmonic systems this is no longer the case, being the spectral weight broadened over several excited states. In particular, for rotating molecules, the anharmonicity of librations makes them hybridize with the vibrons. Therefore, the vibron modes computed by diagonalizing $\tilde{G}_{ij}$ are underestimated by the mixing with much softer librations. 

In our approach, we solve this issue by defining
appropriate displacement operators $\delta\bar{\mathbf{x}}$ to partially decouple vibrational and rotational modes. These operators are effectively restricted to a subspace of the full Fock space. Henceforth, $\delta\bar{\mathbf{x}}$ will be defined as constrained displacements. 
We express them in spherical coordinates around the molecular equilibrium positions and we expand their angular components for small librational angles. We show that in this way both librational and vibron modes are accurately determined by our method even in the rotor limit, when the librational amplitude is maximized.

\subsection{Two-dimensional rotor model}
\label{model}

As a first benchmark application, we take the model of a rotating dimer, described by the two-dimensional (2D) potential that reads as:
\begin{equation}\label{eq:2dpot_main}
	V_c(r,\phi)=d_M \cdot \left[ 1-e^{-a_M\cdot \left( r - r_0 \right)} \right]^2+c \cdot \frac{\text{sin}^2\left( \frac{\phi-\phi_M}{2}\right)}{r^2},
\end{equation}
where $r_0$ is the classical bond length of the dimer, oriented such as to form an angle $\phi_M$ with the $x$ axis. While the vibron mode is controlled by the $d_M$ and $a_M$ parameters, the amplitude of the libration is tuned by the strength $c$ of the external confining angular field (see Fig.~\ref{fig:model_pot}(b) for a plot of the potential as a function of $c$). The other parameters are kept fixed. We set 
$d_M$=0.15 Hartree, $a_M$=1.1 Bohr$^{-1}$ and $r_0$=1.4 Bohr. Then, we numerically solved the corresponding 2D Schr{\"o}dinger equation. We note that the rotational symmetry of the problem allows one to identify two quantum numbers, the radial and the angular one. 
In Fig.~\ref{fig:model_pot}(c), $E_{l,0}$ ($E_{0,l}$) indicates the $l$-th excited state in the vibrational (rotational) channel. 

We first obtain the GEV frequencies by solving Eq.~(\ref{eq:gev}) with the unconstrained displacements $\delta \mathbf{x}=\mathbf{x} - \langle \mathbf{x} \rangle$, where $\mathbf{x}$ is expressed in polar coordinates, such that its components read:
\begin{eqnarray}\label{eq:dxdx_unconstrained}
       x_1 & = & r \ \text{cos}(\phi_M+\Delta\phi), \nonumber \\
        x_2 &= & r \ \text{sin}(\phi_M + \Delta\phi),
\end{eqnarray}
with $\Delta\phi=\phi-\phi_M$. From the upper panel of Fig.~\ref{fig:model_pot}(c), one can immediately see that the GEV eigenvalues of $\tilde{\mathbf{G}}^{-1}$ do not reproduce the exact vibron frequency for small enough $c$.
Indeed, already for a moderate libration regime ($c < 0.2$), the vibron mode is underestimated due to the hybridization with the libration mode, which is instead very well reproduced.
To illustrate the origin of this failure
in a more quantitative way, we study the spectral function of the Green's function in Eq.~(\ref{eq:imag_time_matsu_green}). 
For this simple 2D rotor model, it can be computed exactly, and it is plotted in Fig.~\ref{fig:model_pot}(e) in light green lines. Fig.~\ref{fig:model_pot}(e) shows that, as $c$ gets smaller, more librational excitations become lower in frequency and larger in strength than the first vibron peak, indicated by an arrow in the plot. The vibron peak coincides with the transition from the ground state to the first vibrational level (E$_{1,0}$-E$_{0,0}$). The corresponding vibron eigenvalue 
yielded by 
Eq.~(\ref{eq:gev}) 
is
thus a mix between the physical
vibron mode and the rotational states whose spectral weights are larger than (or comparable with) the vibron and for which E$_{0,J}<$ E$_{1,0}$. This mixing is clearly seen also in the shape of the ground state distribution, shown in Fig.~\ref{fig:model_pot}(d).
On the other hand, in the opposite limit of $c \rightarrow 1$, describing the dimer strongly confined along a preferred direction, one recovers the correct results for the fundamental modes because, in this limit, the first two strongest peaks in the spectral function are those related to $E_{1,0}$ and $E_{0,1}$. Thus,
the mixing between librational and vibrational sectors is suppressed and the result is unbiased.

\begin{figure*}[!t]
	\includegraphics[scale=0.315]{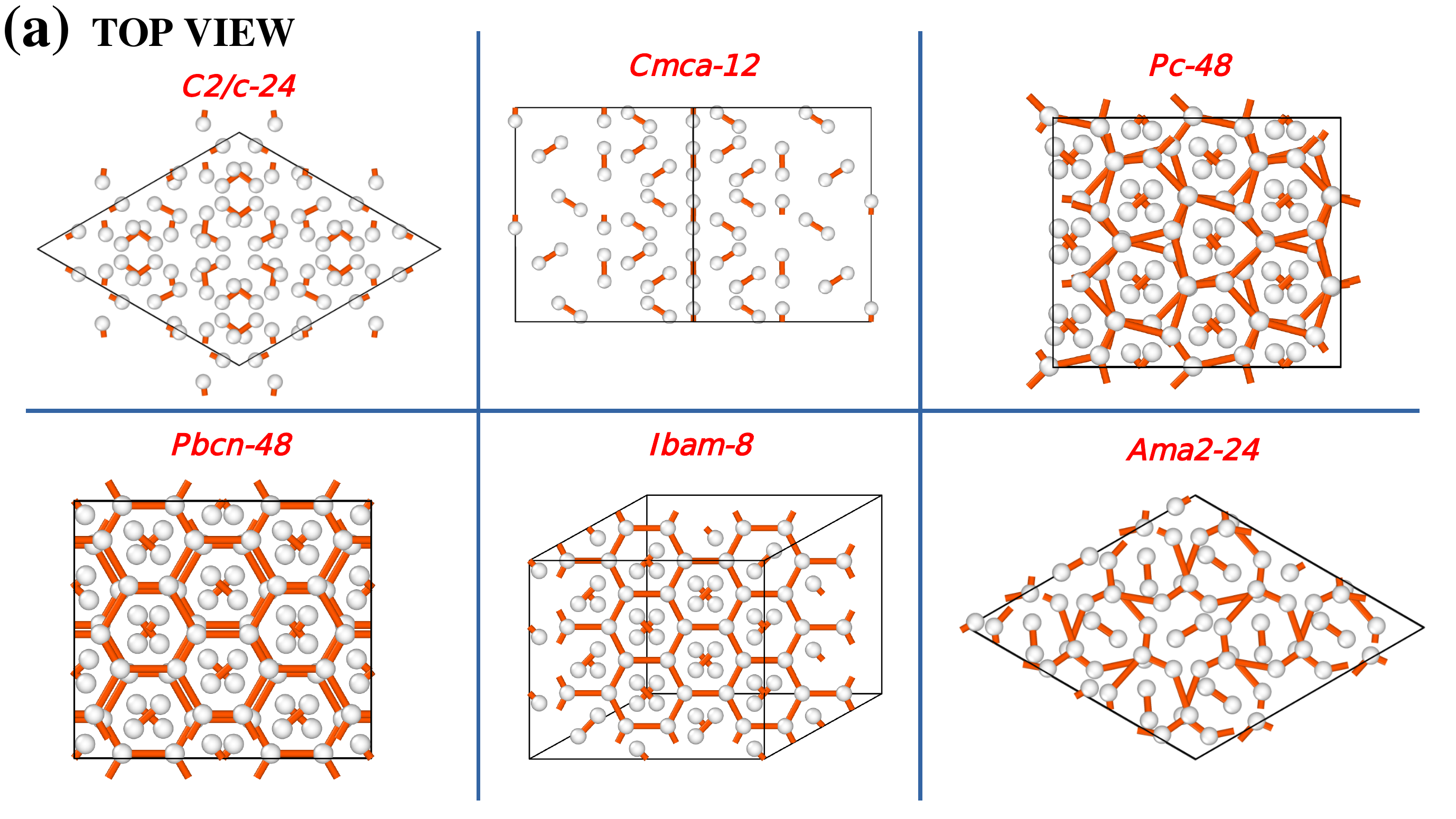}
	\includegraphics[scale=0.315]{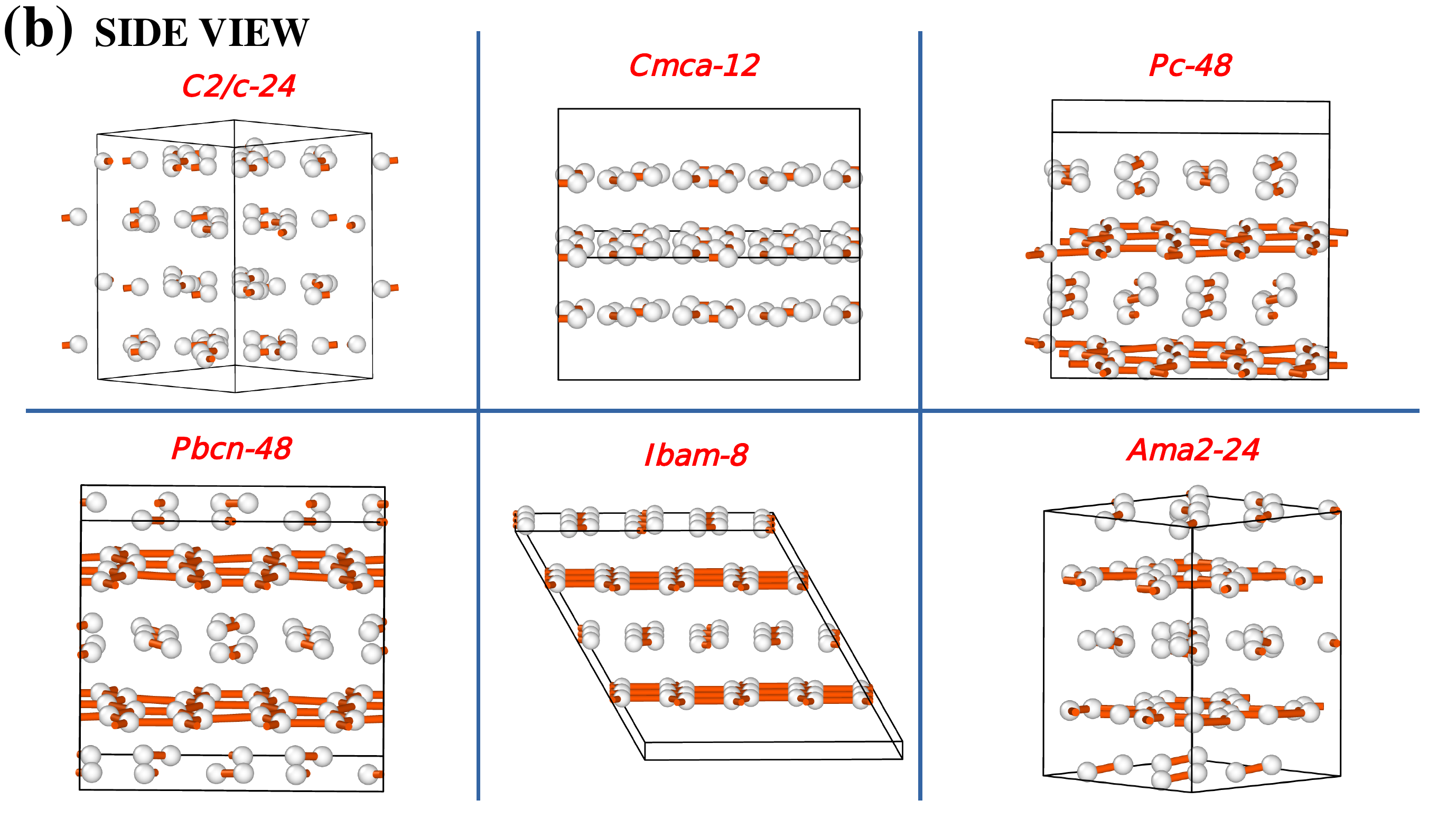}
	\caption{Geometries of the structures analysed in this work. (a) Top view of the supercells. (b) Side view. All the supercells contain 96 hydrogen atoms. The plots are generated with a cutoff for the bonds equal to 2.2 Bohr.}\label{fig:geometries}
\end{figure*}

The failure for small coupling strength $c$ in Eq.~(\ref{eq:2dpot_main}) is fixed by the \emph{constrained} displacement operators
$\delta \bar{\mathbf{x}}=\bar{\mathbf{x}} - \langle \bar{\mathbf{x}} \rangle$, where $\bar{\mathbf{x}}$ in a 2D setting reads:
\begin{eqnarray}\label{eq:dxdx_constrained}
     \bar{x}_1 & = &
r \ \text{cos}(\phi_M) -r \ \text{sin}(\phi_M)\text{sin}(\Delta\phi), \nonumber \\
    \bar{x}_2 &= & 
        r \ \text{sin}(\phi_M) +r \ \text{cos}(\phi_M)\text{sin}(\Delta\phi).
\end{eqnarray}
The displacements $\delta \bar{\mathbf{x}}$ defined by Eq.~(\ref{eq:dxdx_constrained}) have been obtained from Eq.~(\ref{eq:dxdx_unconstrained}) by performing an expansion around the average angle of rotation $\phi_M$ up to the second order in $\Delta\phi$ for the ``slow'' cosine dependence, while keeping the full $\Delta\phi$ dependence in the sine component. This is enough to suppress the hybridization in the relevant matrix elements of Eq.~(\ref{eq:lehmann_green}), and get a very accurate phonon determination for both librational and vibron modes through Eq.(\ref{eq:gev}), as shown in the lower panel of Fig.~\ref{fig:model_pot}(c).
The effect of the \emph{constrained} displacements is apparent in the spectral functions of Fig.~\ref{fig:model_pot}(e) (shown in blue color), where the secondary rotational poles having larger strength than the vibron are suppressed.
This is related to the fact that, by using the \textit{constrained} displacement operators, the matrix elements that couple the vibron mode with these excited rotational levels become negligible.

\subsection{Phase III}
\label{phase_III}
After the 2D rotor model, we benchmark our method in a realistic setting, by investigating the vibron modes of phase III of solid hydrogen. In the three-dimensional molecular case, the strategy to constrain the displacement operators is described by the following steps:
\begin{enumerate}
    \item we read the atomic Cartesian coordinates 
    from the output trajectories of the PIMD simulations;
    \item once the molecules are identified in the simulation cell, we express their positions in spherical coordinates;
    \item we go back from spherical to Cartesian coordinates, by defining the new constrained positions for each molecule in the following way:
    \begin{equation}
        \bar{x}_i^{(\pm)} = x_{c.m.,i} \pm \frac{\bar{x}_{R,i}}{2}, 
        \label{eq:polar_to_cartesian}
    \end{equation}
    where in the above notation ``$\pm$'' identifies the two atoms of the dimer, $c.m.$ is its center of mass and $\bar{x}_{R,i}$ are obtained from the constraining relations:
    \begin{equation}\label{eq:kubo_mod_3d}
    \begin{split}
    x_{R,1} &= r\cdot \text{sin}(\theta) \text{cos}(\phi)\\
            & \eqsim r \cdot  \left(\text{sin}(\theta_M)+\text{cos}(\theta_M)\text{sin}(\Delta\theta) \right)\cdot \\ & \ \ \ \ \ \ \ \cdot \left(\text{cos}(\phi_M)-\text{sin}(\phi_M)\text{sin}(\Delta\phi) \right) \equiv \bar{x}_{R,1}, \\
    x_{R,2} &= r\cdot \text{sin}(\theta) \text{sin}(\phi)\\
            & \eqsim r \cdot \left(\text{sin}(\theta_M)+\text{cos}(\theta_M)\text{sin}(\Delta\theta) \right)\cdot \\ & \ \ \ \ \ \ \ \cdot \left(\text{sin}(\phi_M)+\text{cos}(\phi_M)\text{sin}(\Delta\phi) \right) \equiv \bar{x}_{R,2}, \\
    x_{R,3} &= r \cdot \text{cos}(\theta) \\
            & \eqsim r \cdot \left(\text{cos}(\theta_M)-\text{sin}(\theta_M)\text{sin}(\Delta\theta) \right) \equiv \bar{x}_{R,3}, \\
    \end{split}
    \end{equation}
    where $\Delta\theta=\theta-\theta_M$, $\Delta\phi=\phi-\phi_M$, being $\theta_M$ and $\phi_M$ the polar and azimuthal angles of the molecular axis at equilibrium.
\end{enumerate}
Due to the nonlinear relations between Cartesian and spherical coordinates, this operation should be performed explicitly over all the beads, and not on the PIMD centroids. Therefore, the PIMD displacement operator is then computed as $\delta \bar{x}_i = \frac{1}{P} \sum_{j=1}^P \bar{x}_i^{(j)} - \langle \bar{x}_{i} \rangle$, where $P$ is the number of beads in the PIMD simulation~\cite{Mouhat_2017}.
Although the constraints in Eqs.~(\ref{eq:polar_to_cartesian}) and (\ref{eq:kubo_mod_3d}) are defined for molecular dimers, it is worth to note that this procedure can be easily generalized to an arbitrary structure. 
Indeed, the constraints can be implemented in any situation once the preferential vibrational axis, around which the slow cosine-dependence is expanded, has been chosen. 
In the dimer case, this choice is dictated by the direction of the bond linking the two atoms forming the dimer. In the most general case instead, one should choose that axis according to the preferred 
symmetries of the system.
\begin{figure*}[!t]
	\includegraphics[scale=0.9]{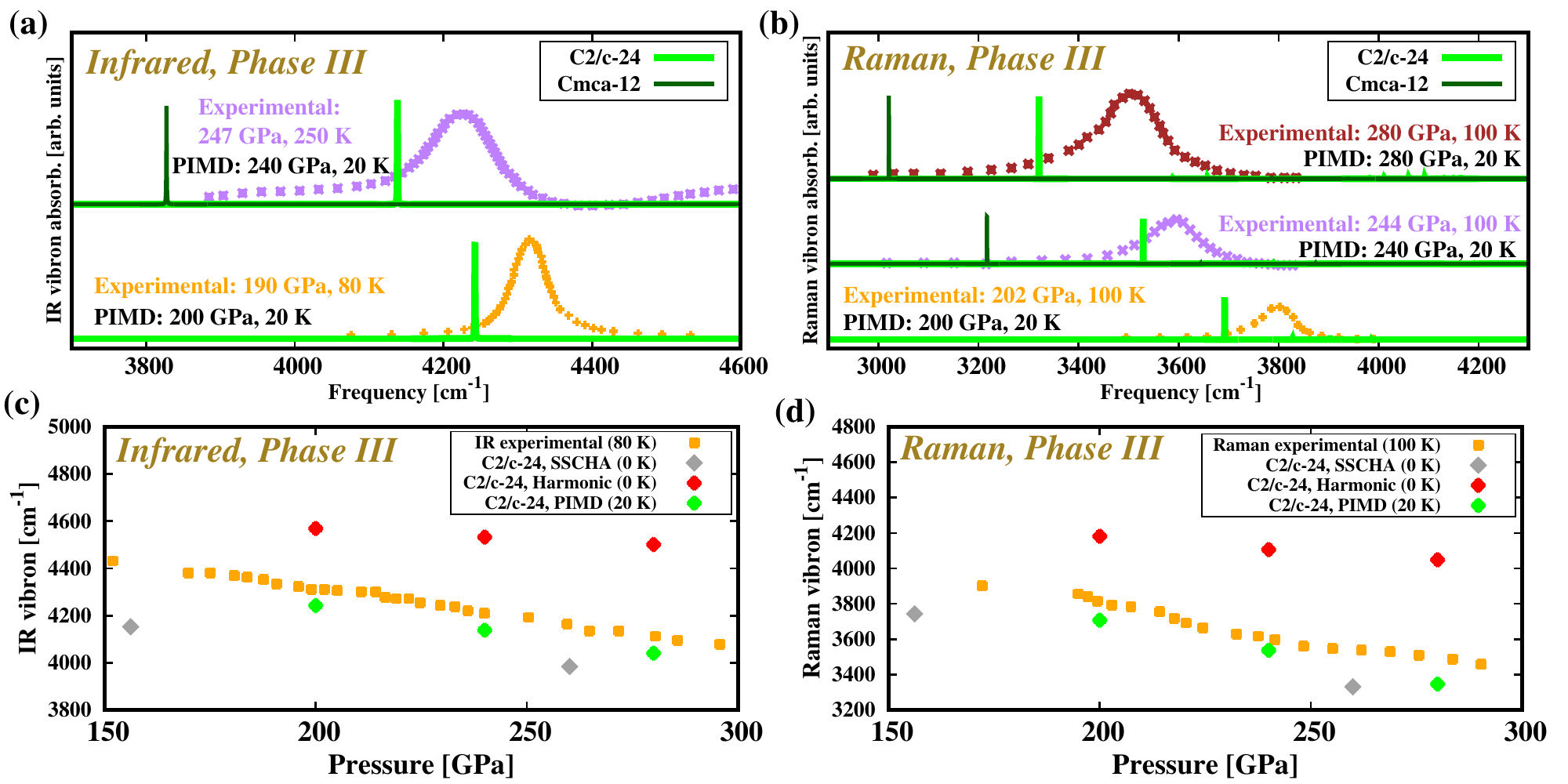}
	\caption{Experimental measurements vs theoretical predictions for phase III. (a) Experimental IR vibron peaks, taken from~\cite{Loubeyre_2020} at 190 GPa and from~\cite{Zha_2012} at 247 GPa, versus theoretical IR anharmonic peaks of C2/c-24 (200 and 240 GPa) and Cmca-12 (240 GPa) geometries. (b) Experimental Raman vibron peaks, taken from~\cite{Eremets_2019}, compared with theoretical Raman anharmonic peaks of C2/c-24 (200, 240 and 280 GPa) and Cmca-12 geometries (240 and 280 GPa). (c) Experimental infrared~\cite{Loubeyre_2020} and (d) Raman~\cite{Eremets_2019} vibron peaks as function of pressure compared with theoretical predictions of the C2/c-24 geometry at the harmonic level (red), SSCHA (grey) and PIMD (green). SSCHA values are taken from~\cite{Monacelli_2021}. Here and in the following figures, literature data are digitalized using WebPlotDigitizer~\cite{Rohatgi_2020}.}\label{fig:phase_3}
\end{figure*}

Phase III is stable 
in a range of pressures between
160 GPa and 420 GPa at temperatures lower than 200 K~\cite{Loubeyre_2020}. 
The well established symmetry for phase III is the C2/c-24 geometry~\cite{Pickard_2007,Drummond_2015,Monacelli_2021,Monacelli_2022} (see Fig.~\ref{fig:geometries}). 
Thus, we perform three different simulations for the C2/c-24 at 200, 240 and 280 GPa, by running PIMD with BLYP-driven forces. The BLYP functional~\cite{Lee_1988,Becke_1988} has already been proved to give results in reasonable agreement with diffusion Monte Carlo calculations in this range of pressures~\cite{Drummond_2015,Azadi_2017}. For testing purposes, we examine also the Cmca-12 competing configuration~\cite{Singh_2014,Drummond_2015} at 240 and 280 GPa.  
Computed vibron 
spectra with full inclusion of NQE
are shown in Figs.~\ref{fig:phase_3}(a) and \ref{fig:phase_3}(b), where we compare the anharmonic IR and Raman peaks with experiments. In this comparison, we weighted
the vibron modes using Born effective charges and Raman tensor to reproduce the IR and Raman intensities, respectively.
Theoretical peaks are delta-like because, by means of our approach, we have access to the static self-energy, without lifetime effects. 
We observe a nice agreement between the C2/c-24 IR and Raman peaks and the experimental ones, even if 
the frequencies of the computed spectra
are still underestimated by $\sim$100 cm$^{-1}$.
Furthermore, a large frequency difference is found between the C2/c-24 and Cmca-12 geometries, with the latter in clear disagreement with the experiments. This benchmark case shows that our PIMD phonons approach has the predictive power to attribute the right symmetry to a given hydrogen phase, from a genuine comparison with the experimental vibrons.

In Figs.~\ref{fig:phase_3}(c) and \ref{fig:phase_3}(d) we report the pressure behaviour of the IR and Raman experimental peaks, respectively, and we compare them with the C2/c-24 peaks evaluated at different levels of theory.
We observe a strong renormalization of the anharmonic PIMD vibron frequencies with respect to the harmonic case. Also, the vibron energy slope is correctly 
reproduced only when NQE are taken into account. Furthermore, we find that our PIMD vibron frequencies are closer to the experiment than the ones computed by the time dependent SSCHA
framework using the same BLYP functional~\cite{Monacelli_2021}. 
This is due to the SSCHA Gaussian ansatz limitations in presence of strong librational modes, and it is rationalized by a detailed analysis of the 2D dimer model using the SCHA framework (see Supplementary Information~\cite{supp}).

\subsection{Phase IV}
\label{phase_IV}
We then addressed the open problem of determining the right symmetry for phase IV~\cite{Zha_2012,Zha_2013,Loubeyre_2013,Goncharov_2019}. Experimentally, it is stable above 300 K and 220 GPa~\cite{Howie_2012,Howie_2012_2}. 
A number of theoretical works~\cite{Liu_2012,Pickard_2012,Goncharov_2013,Azadi_2018,Li_2020} led to a list of possible candidates for its structural characterization. 
\begin{figure*}[!t]
	\includegraphics[scale=0.83]{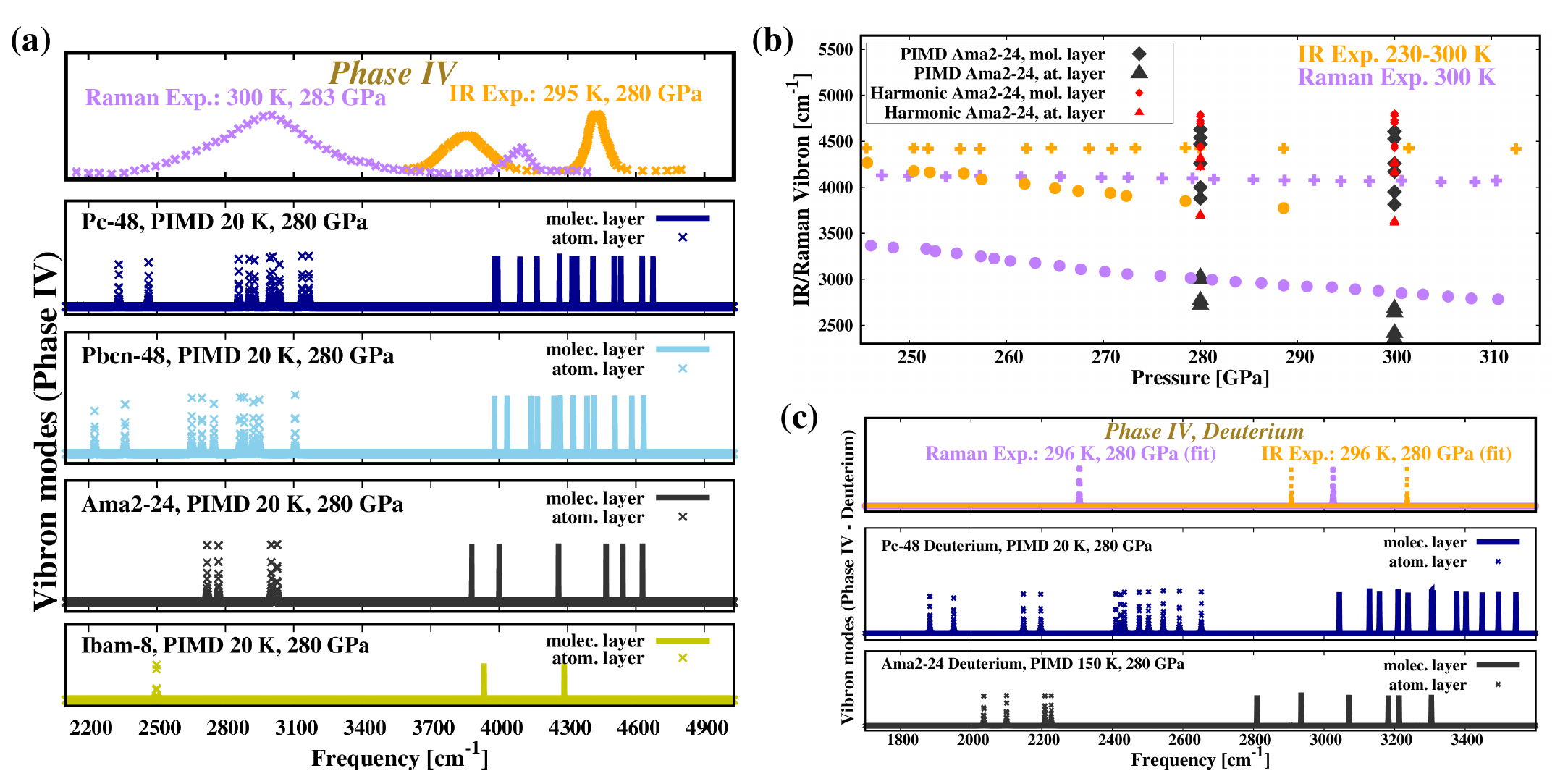}
	\caption{Experimental measurements vs theoretical predictions for phase IV. (a) Top panel: Experimental IR and Raman vibron peaks for phase IV at 280 GPa. Raman experimental data are taken from~\cite{Howie_2012} while infrared experimental from~\cite{Zha_2013}. Lower panels: PIMD vibron modes for the four different geometries: Pc-48; Pbcn-48; Ama2-24; Ibam-8. Continuous lines represent modes projected over the molecular layers, while points are projections over atomic-like layers. (b) Experimental infrared~\cite{Zha_2013} and Raman~\cite{Howie_2012} vibron peaks as function of pressure compared with theoretical predictions of the Ama2-24 geometry at the harmonic level (red) and PIMD (dark grey). (c) Experimental IR and Raman vibron peaks for deuterium phase IV at 280 GPa, extrapolated by fitting the slope of the vibron energy as a function of pressure. Raman experimental data of Raman vibron are taken from~\cite{Howie_2012} while infrared experimental from~\cite{Loubeyre_2013}. Lower panels: anharmonic theoretical vibron modes for Pc-48 and Ama2-24.}\label{fig:phase_4}
\end{figure*}
Here, we consider four different geometries, 
already introduced in previous works, that are all characterized by alternating layers of strongly bonded H$_2$ molecules and atomic-like layers (G-layers): 
Pc-48~\cite{Pickard_2012}, having a strongly distorted honeycomb lattice as G-layer; Pbcn-48~\cite{Pickard_2007}, where the honeycomb lattice is less distorted and slightly dimerized; Ama2-24~\cite{Li_2020}, where the G-layer has a typical C2/c arrangement; Ibam-8~\cite{Pickard_2007} with a perfectly hexagonal G-layer. 
See Fig.~\ref{fig:geometries} for their graphical representation.
From previous PIMD results~\cite{Rillo_2018}, confirmed by our simulations, it turns out that these structures are clearly quantum crystals~\cite{Cazorla_2017}, with
spatial fluctuations even larger than in phase III. 
At variance with phase III, where the H$_2$ molecules librate but never perform full $\pi$-rotations, the molecules belonging to the molecular layers of phase IV are free to rotate around their center of mass. They behave very similarly for all four geometries, with an average bond length of 1.36 Bohr in every case (see Tab.~\ref{tab:lengths} in \cite{supp}). 
Moreover, high temperatures lead to an additional drift of their center of mass. Therefore, we study their vibron modes at 280 GPa and 20 K, thermodynamic conditions that allowed us to compute less noisy PIMD phonons. Nevertheless, quantum fluctuations are strong enough to place these structures in a regime where IR and Raman matrix elements are no longer reliable within a perturbative approach.
Thus, for phase IV we do not weight the PIMD vibron modes by these matrix elements. We show them all in the lower panels of Fig.~\ref{fig:phase_4}(a) for the four considered geometries and we finally compare their frequencies with corresponding experimental IR and Raman values (top panel of Fig.~\ref{fig:phase_4}(a)).

\section{Discussion}
Experimentally, phase IV is believed to be made of layers of strongly bonded molecules alternated by graphene-like sheets~\cite{Pickard_2012}. A signature of this structure would be a strong softening of the G-layer vibron with increasing pressure. 
Therefore, in order to study the character of the vibron modes of the simulated structures, we project these modes over both molecular and atomic-like layers for each system. 
A striking feature is the clear separation between the vibron frequencies belonging to the molecular layers from the atomic ones. Interestingly enough, quantum anharmonicity not only renormalizes the full spectrum, but also widens the 
gap between these two sets of vibrons with respect to the harmonic approximation. 
This effect is present in all geometries analyzed, being the mildest in Ibam-8.
From this analysis, we deduce that both IR peaks measured in experiments and reported in the top panel of Fig.~\ref{fig:phase_4}(a) should necessarily be localized over molecular layers. This is at variance with a former interpretation, which assigned the softer IR peak to atomic-layer modes~\cite{Loubeyre_2013}. The rationale behind it is that strong NQE smooth out the structural differences between competing arrangements, particularly in the atomic layer, and make it closer to an undistorted hexagonal shape (see Fig.~\ref{fig:quantum_pict} in \cite{supp}). This is in accordance to the fact that, for a perfect graphene-like arrangement, the IR response is absent.
From Fig.~\ref{fig:phase_4}(a), by comparing
the predicted vibron frequency distribution 
with the experimental one, it turns out that Ama2-24 is the best-matching geometry for phase IV. 
\begin{table}
\caption{PIMD expectation value of the Hamiltonian ($\langle H \rangle$) vs static DFT energies. 
The energies with quantum nuclei are computed using the virial estimator~\cite{Mouhat_2017} for the kinetic contribution. 
All the PIMD simulations are performed at 20 K and using 120 beads. Energies are given in eV and are normalized by the number of atoms in the simulation cells.}
\begin{ruledtabular}
\begin{tabular*}{\textwidth}{{cccc}}\label{tab:energies}
         {\small \textit{Structure}} & {\small \textit{Pressure (GPa)}} & {\small \textit{$\langle H \rangle$ PIMD}} & {\small \textit{static DFT}} \\
\hline
 $C2/c$--$24$ & 200 & -14.2816(1) & -14.5088  \\ 
 $C2/c$--$24$ & 240 & -14.0890(2) & -14.3254 \\ 
 $C2/c$--$24$ & 280 & -13.9095(1) & -14.1535 \\ \\
 $Cmca$--$12$ & 240 & -14.0672(1) & -14.3091  \\ 
 $Cmca$--$12$ & 280 & -13.8912(2) & -14.1398  \\ \\
 $Pbcn$--$48$ & 280 & -13.9054(2) &  -14.1467 \\ \\ 
 $Pc$--$48$ & 280 & -13.9089(2) &  -14.1504 \\ \\
 $Ibam$--$8$ & 280 & -13.8740(6) & -14.1070 \\ \\
 $Ama2$--$24$ & 280 & -13.9103(6) & -14.1512  \\ \\
\end{tabular*}
\end{ruledtabular}
\end{table}
A further verification of the symmetry attribution for phase IV comes from studying the static and fully anharmonic lattice energies in Tab.~\ref{tab:energies}. Indeed, for a pressure equal to 280 GPa, Ama2-24 is the most competitive among phase IV candidates both at the static and anharmonic lattice level, followed by Pc-48. 
We notice however that, from a structural perspective, NQEs make the difference between these various symmetries much milder than the ones present in static configurations.

Based on the above analysis,
we chose the Ama2-24 geometry to perform another simulation at different pressure (300 GPa) to inspect if, like for phase III, the pressure dependence
follows the slope of the experimental vibron energies. We show the results in Fig.~\ref{fig:phase_4}(b), where we compare the harmonic and PIMD vibron frequencies (in red and black colours, respectively) with the experimental values. We observe that the lowest Raman peak is caught only by PIMD frequencies and that its slope is correctly described by anharmonic vibron modes of atomic character. All PIMD vibron frequencies above are of molecular character, and they are spread over an energy interval compatible with the experimental findings over the full pressure range analyzed here.

To confirm the scenario, we also performed a simulation of Ama2-24 and Pc-48 using deuterium in place of hydrogen, at 280 GPa. At this pressure and with the deuterium mass, both symmetries have the tendency to transform into the C2/c-24 geometry for low enough temperatures. Ama2-24 is stable at 150 K, while at 20 K we have been able to run Pc-48 simulations for 4.5 ps, a time sufficiently long to estimate the PIMD phonons. As shown in Fig.~\ref{fig:phase_4}(c), also in this case Ama2-24 seems to best fit the experiments. The underestimation of the softer Raman peak predicted in Ama2-24 is consistent with what found in C2/c-24 for phase III.

In summary, in this work we introduced an accurate method for computing phonons of molecular solids strongly affected by NQE. The framework proposed here paves the way towards a robust assessment of crystalline symmetries, even in cases where structural information comes primarily from vibrational spectroscopy. This is particularly relevant in quantum crystals, where the usual harmonic approximation badly fails, preventing a direct comparison with experiments. 
The present approach is suitable to better quantify their unique properties, such as polaron effects and superconductivity in hydrogen-rich materials. 

\section{Methods}
The PIMD simulations are carried out at 20 K using 120 beads to take into account quantum effects. Nuclei are evolved in time using the PIOUD integrator~\cite{Mouhat_2017} with a time step equal to 0.5 fs and a friction parameter of the Langevin thermostat equal to 1.46$\cdot$10$^{-3}$ atomic units. The latter value is the same as in Ref.~\cite{Mouhat_2017}, where it is found to be optimal for both stochastic and deterministic forces. Simulations lasted around 10 ps, until the convergence on vibron modes at Gamma is reached. Forces are computed from the Born Oppenheimer potential energy surface (PES) evaluated at Density Functional Theory level within the Quantum Espresso~\cite{qe1} engine. In particular we used a BLYP 
functional for computing the PES and an LDA functional for the evaluation of the Born effective charges and Raman tensors. The choice of the BLYP functional comes from the benchmark tests using Quantum Monte Carlo calculations performed in Ref.~\cite{Clay_2014} and also from the fact that, between the different available functionals, geometries computed at BLYP level give the lowest Diffusion Monte Carlo energies \cite{Azadi_2017,Drummond_2015}. The wavefunction cut-off for the PES is set to 60 Ry (300 Ry for the charge density), while the Fermi smearing is Gaussian and set equal to 0.03 Ry. PIMD simulations are performed using supercells containing in each case 96 hydrogen atoms and the corresponding supercell reciprocal space mesh is always equal to 4x4x4.
For phase III, Raman active modes are evaluated from the response function in the following way~\cite{Monacelli_2021}:
\begin{equation}\label{eq:raman_expr}
    I_{Raman}(\omega) \propto - \sum_{\alpha=1}^3 \sum_{a,b} \frac{A_{\alpha \alpha,a}A_{\alpha \alpha,b}}{\sqrt{m_a m_b}} \ \text{Im}\left[ G_{a,b}(\omega) \right],
\end{equation}
where $a,b$ run over all the 3N degrees of freedom, $A$ is the Raman tensor that we derived from static DFT calculations using the Quantum Espresso suite~\cite{qe1}, $m_i$ is the mass of the i-th degree of freedom, Im is the imaginary part, $G^{-1}_{a,b}(\omega)= \omega^2 - \Phi_{a,b}$ and $\Phi_{a,b}$ is the force constant matrix at $\Gamma$ that we build from PIMD simulations, therefore already including the renormalization due to anharmonic effects.\\
Infrared active modes are instead computed using the Born effective charge tensors $Z^*$, with the expression:
\begin{equation}\label{eq:ir_expr}
    I_{IR}(\omega) = \sum_{\alpha,\beta=1}^3 \sum_{a,b} \frac{Z^*_{\alpha,a}Z^*_{\beta,b}}{\sqrt{m_a m_b}} \ \text{Im}\left[ G_{a,b}(\omega) \right].
\end{equation}
In particular, the points reported in Fig.~\ref{fig:phase_3} are computed by averaging the frequencies with the spectra obtained from Eqs.~(\ref{eq:raman_expr}) and (\ref{eq:ir_expr}), i.e:
\begin{equation}
    \begin{split}
    & \omega_{Raman,IR} = \frac{\int d \omega \ \omega \cdot I_{Raman,IR}(\omega) }{\int d \omega \ I_{Raman,IR}(\omega)}.
    \end{split}
\end{equation}

For phase IV, we find that the above methods to estimate IR and Raman tensors fail because the system is strongly anharmonic and the small displacement around equilibrium positions employed to estimate tensors are no longer reliable.
In Figs.~\ref{fig:phase_4}, phonons at $\Gamma$-point of the Brillouin zone are projected over the layer of strongly bonded molecules (solid lines) and over the atomic-like layer (points). The projector operator is built by orthonormalizing the N$_P$, where N$_P$ is the number of projections, 3N-dimensional vectors $|v^{proj} \rangle$, representing the atoms of each H$_2$ molecule over which we project. 
Then, the reduced eigenvectors are found by:
\begin{equation}
    \phi^{proj}_{j,a} = \sum_{b} |v^{proj}_{j,b} \rangle \langle v^{proj}_{j,b} | \Phi_{a,b} \rangle.
\end{equation}
The projected density of states (DOS) is finally obtained using the following expression:
\begin{equation}
    PDOS (\omega) = \sum_{a=1}^{3N} \langle \phi^{proj}_a | \phi^{proj}_a \rangle \delta(\omega - \bar{\omega}_{a}).
\end{equation}\\

\section*{Acknowledgements}
T.M. and M.C. thank the French grand {\'e}quipement national de calcul intensif (GENCI) for the computational time provided under Project No. 0906493.
All authors thank Lorenzo Paulatto, Lorenzo Monacelli and Michele Lazzeri for useful discussions.
\section*{Funding}
This work is supported by the European Centre of Excellence in Exascale Computing TREX (Targeting Real Chemical Accuracy at the Exascale). This project has received funding from the European Union's Horizon 2020 (Research and Innovation program) under grant agreement no. 952165.
\section*{Author Contributions}
TM coded the numerical algorithms and performed the calculations. 
All authors developed the theory and analysed the data.
TM
wrote the paper with
contributions
from MC and RV.  MC proposed and led the project. 
\section*{Competing Interests}
The authors declare no competing interests.
\section*{Data Availability}
The data that support the findings of this study are available from the corresponding author upon reasonable request.
\section*{Code Availability}
The codes implementing the calculations of this study are available from the corresponding author upon request.

\bibliography{biblio}

\clearpage

\onecolumngrid
\section*{\LARGE SUPPLEMENTARY INFORMATION}
\subsection*{Tommaso Morresi,$^{1\ast}$ 
Rodolphe Vuilleumier,$^{2}$
Michele Casula$^{1\ast}$}
\noindent
$^{1}$Institut de Min{\'e}ralogie, de Physique des Mat{\'e}riaux et de Cosmochimie (IMPMC), Sorbonne Universit{\'e}, CNRS UMR 7590,  MNHN, 4 Place Jussieu, 75252 Paris, France\\
$^{2}$PASTEUR, D{\'e}partement de chimie, {\'E}cole normale sup{\'e}rieure, PSL University, Sorbonne Universit{\'e}, CNRS, 75005 Paris, France\\~\\

\twocolumngrid

\subsection*{Two-dimensional roto-vibrational model}
The shape of the potential in Eq.~(\ref{eq:2dpot_main}) for three different values of $c$ is presented in Fig.~\ref{fig:model_pot}(b). 
In particular, the limit $c \to 0$ (c$\sim$0.005) corresponds to the case in which the dimer is free to rotate, while the large $c$ limit (c$\sim$0.5) describes the molecule confined along one preferred direction. In the intermediate range (c$\sim$0.05), the dimer cannot perform a full $\pi$-rotation but the amplitude of the librational motion is more or less wide depending on the value taken by the $c$ parameter. 
As we have done in Ref.~\cite{Morresi_2021}, we carry out the analysis by solving numerically the Schr{\"o}dinger equation at low temperature (20 K) to minimize thermal effects and to focus purely on quantum effects. The details of the numerical scheme can be found in Appendix E of Ref.~\cite{Morresi_2021}. 
\begin{figure}[!h]
	\includegraphics[scale=0.68]{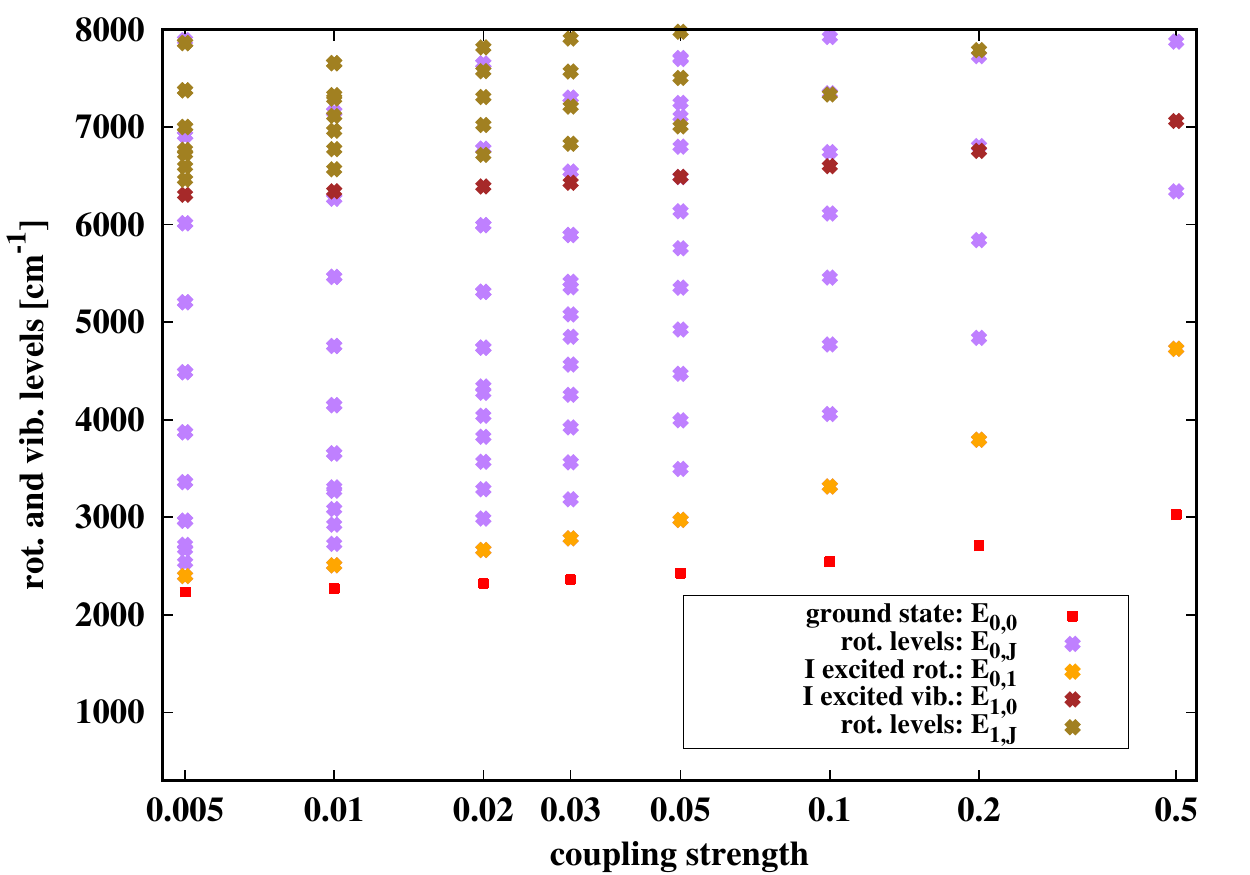}
	\caption{Eigenvalues of the Schr{\"o}dinger equation for the potential in Eq.~(\ref{eq:2dpot_main}). In particular we highlight the ground state level (red), the first excited state (orange) which is a rotational state in the range of $c$ considered here, the first vibrational excited state (brown) and all the remaining states which have rotational nature (purple and olive).}\label{fig:rot_states}
\end{figure}
As seen in the main text (top panel Fig.~\ref{fig:model_pot}(c)), using the \emph{unconstrained} displacement operators one is able to reproduce very well the rotational frequency (orange line in Fig.~\ref{fig:model_pot}(c)). However, for $c < $ 0.2 the computed vibron frequency (brown line in Fig.~\ref{fig:model_pot}(c)) is biased. 
Based on the eigenstates decomposition 
(Eq.~(\ref{eq:lehmann_green}) of the main text and Eq.~(26) of Ref.~\cite{Morresi_2021}), another explanation for this failure can be given by looking at Figs.~\ref{fig:rot_states}, 
reporting the eigenvalues of the system. 
\begin{figure}
	\includegraphics[scale=0.31]{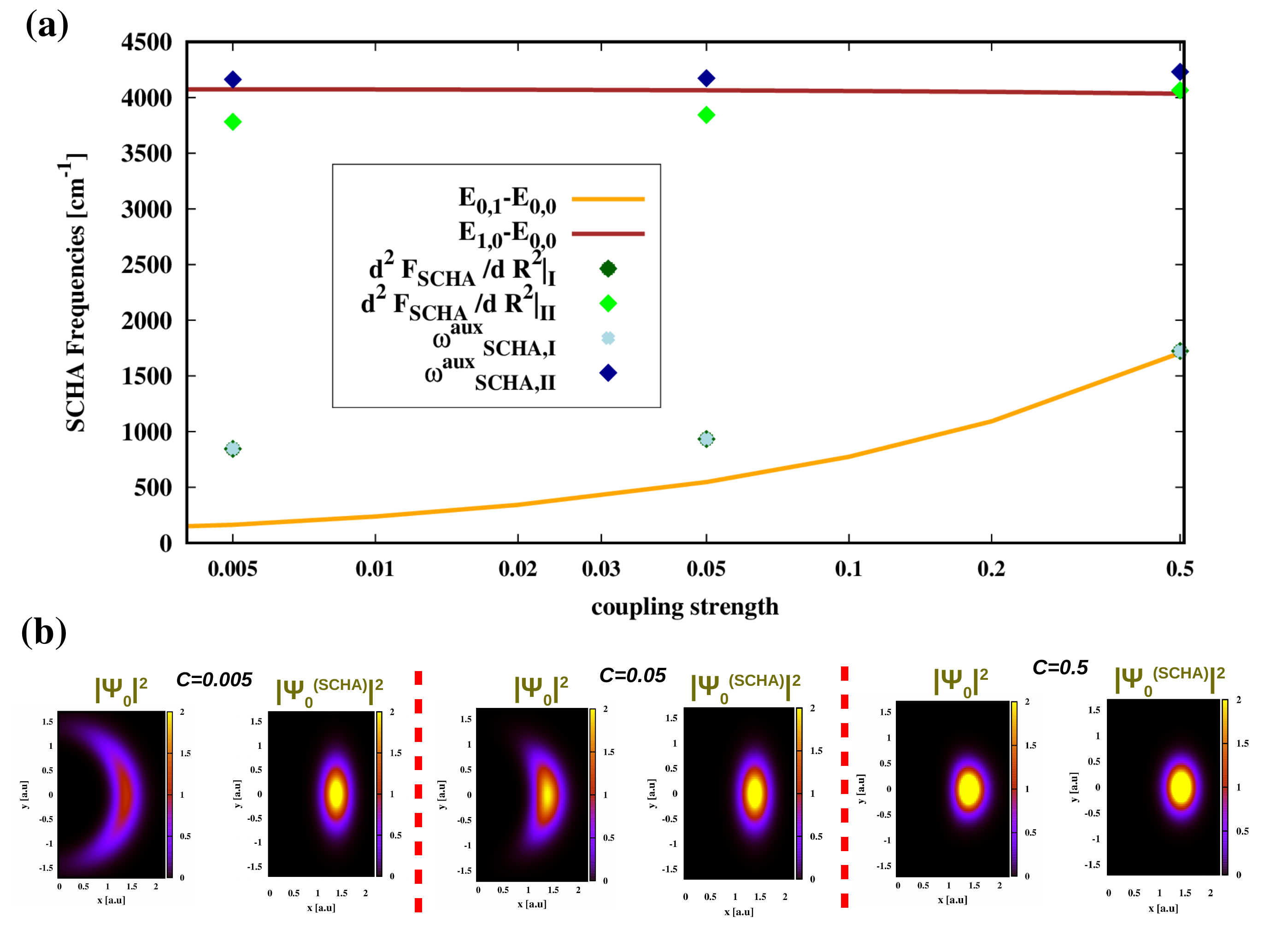}
	\caption{(a) SCHA eigenvalues obtained with the auxiliary hessian matrix and the second derivative of the SCHA free energy for different values of $c$ in Eq.~(\ref{eq:2dpot_main}). (b) Density plot of the square modulus of the ground state wavefunction ($|\Psi_0|^2$) versus SCHA wavefunction ($|\Psi^{\text{(SCHA)}}_0|^2$) for the points reported in panel (a).}\label{fig:scha_model}
\end{figure}
Indeed, while one would expect to find the transition between the first rotational state and the ground state (E$_{0,1}$ - E$_{0,0}$) and the transition between the first vibrational excited state and the ground state (E$_{1,0}$ - E$_{0,0}$) as fundamental modes, the presence of rotational states 
in between for $c < $0.2 clearly biases the evaluation of the vibron mode (i.e., E$_{1,0}$ - E$_{0,0}$). By inspecting the matrix elements coupling different 
excitations, one can see that, even if the transition from state $(1,0)$ to the state $(0,0)$ has the largest weight, the matrix elements coupling higher rotational states (namely, states $(0,J)$ with $J>1$) with the ground state yield a spurious contribution to the calculated vibron frequency 
for $c<$0.2. 

Finally, we would like to compare our framework with the Self Consistent Harmonic Approximation (SCHA)~\cite{Errea_2013,Bianco_2017} results on the very same dimer model.
Indeed, the stochastic SCHA approach provides the main theoretical reference for phase III, beside our calculations (Figs.~\ref{fig:phase_3}(c) and ~\ref{fig:phase_3}(d)). This approach is based on a variational principle for the free energy \cite{Bianco_2017} and includes NQE as well as PIMD. In Fig.~\ref{fig:scha_model} we report the behaviour of the SCHA eigenvalues for the auxiliary Hessian matrix ($\omega^{aux}_{SCHA,I}$ and $\omega^{aux}_{SCHA,II}$) and for the second derivative of the SCHA free energy ($d^2F_{SCHA}/dR^2|_I$ and $d^2F_{SCHA}/dR^2|_{II}$) for a few $c$ values in Eq.~(\ref{eq:2dpot_main}).
In Fig.~\ref{fig:scha_model}(a) we can observe a small underestimation of the vibron mode using the SCHA free energy's second derivative in the region $c<$0.5, while for $c=$0.5 the vibron eigenvalue is correctly reproduced. 
This behaviour is understood by the SCHA Gaussian ansatz wavefunction, that cannot recover the full rotational character of the true ground state wave function for small coupling parameters. This can be clearly seen in Fig.~\ref{fig:scha_model}(b). On the other hand, in the limit of the molecule strongly confined along one direction ($c\geq$0.5 in our model), the SCHA wavefunction reproduces (in a variational sense) very well the true ground state wavefunction and the eigenvalues are accurately estimated. 

\subsection*{Details of the atomic structures}
In Fig.~\ref{fig:geometries} of the main text, we report a picture of the initial supercell configurations for all the geometries studied in this work. 
\begin{figure}[h!]
	\includegraphics[scale=0.31]{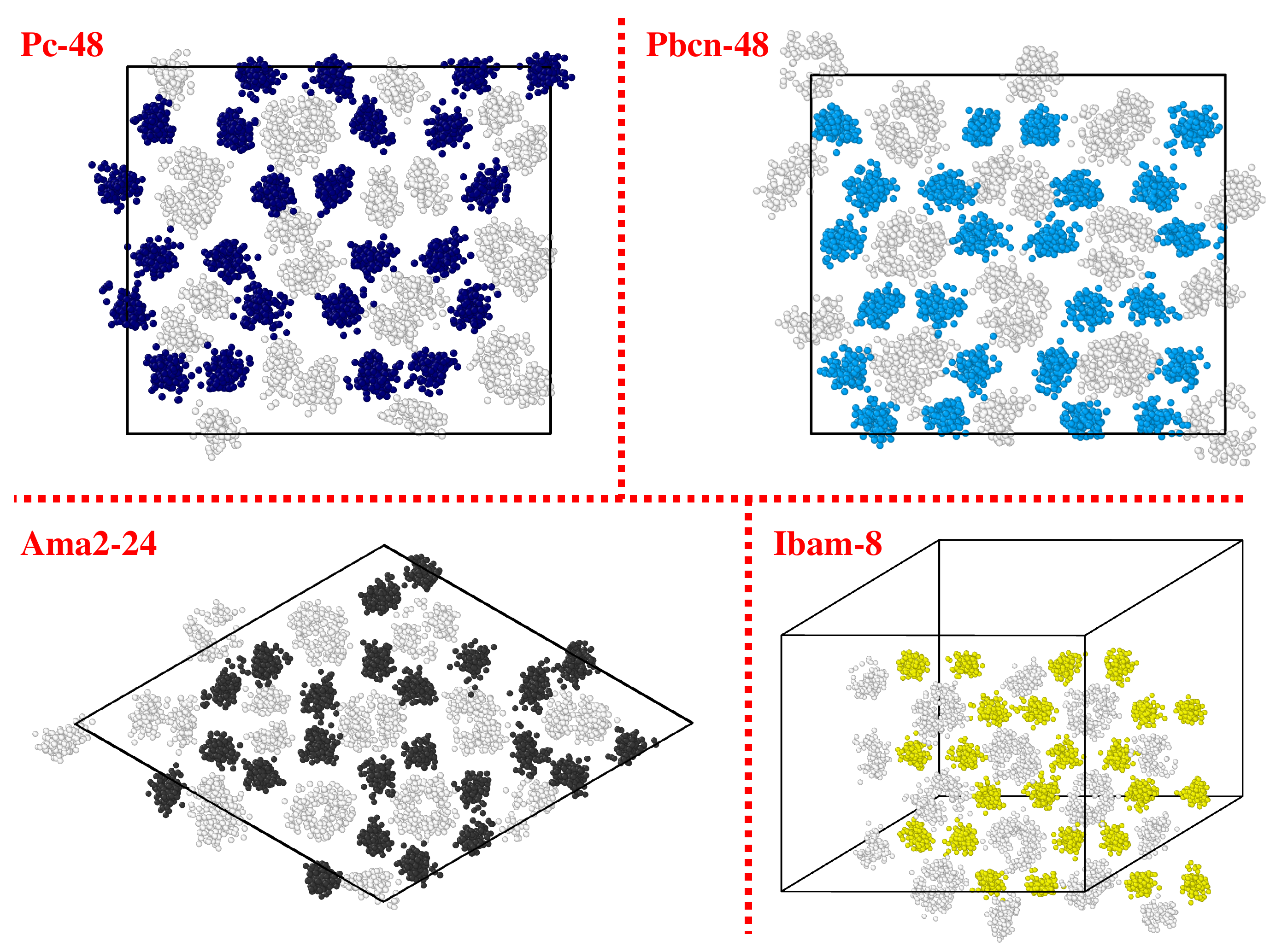}
	\caption{Top view snapshots of two different layers of mixed structures. Top left: Pc-48. Top right: Pbcn-48. Bottom left: Ama2-24. Bottom right: Ibam-8. The colours employed to draw the highlighted G-layers are the same of Fig.~\ref{fig:phase_4}(a) of the main text. Apart from the different lattice symmetries, by plotting all the beads an overall similarity between the geometries can be recognized.}\label{fig:quantum_pict}
\end{figure}
The details of their unit cells are given in the following. 
Purely molecular structures (phase III):
\begin{itemize}
\item \textit{C2/c-24}: base-centered monoclinic structure and four different layers, arranged in an ABCD fashion, of H$_2$ molecules ordered in distorted hexagonal rings;
\item \textit{Cmca-12}: base-centered orthorhombic lattice and two different layers of H$_2$ molecules arranged in an AB fashion;
\end{itemize}
Mixed structures (phase IV):
\begin{itemize}
\item \textit{Pc-48}: simple orthorombic lattice, four layers stacked in an ABAB fashion and the G-layers made by strongly distorted hexagons (corresponding to weak bonded molecules)~\cite{Pickard_2012};
\item \textit{Pbcn-48}: simple orthorombic lattice, four layers stacked in an ABAB fashion and the G-layers made by slightly distorted hexagons (corresponding to very weak bonded molecules)~\cite{Pickard_2007};
\item \textit{Ama2-24}: base-centered orthorombic lattice, four layers stacked in an ABCD fashion and the G-layers very similar to the C2/c layers~\cite{Li_2020};
\item \textit{Ibam-8}: body-centered orthorhombic lattice, four different layers stacked in an ABAB fashion and perfectly hexagonal G-layers~\cite{Pickard_2007}.
\end{itemize}
Finally, in Fig.~\ref{fig:quantum_pict} we exhibit a top-view plot of the mixed structures, with all beads gathered together at given time (that we choose in the middle of the trajectory) of the PIMD simulation. We highlight the G-layer atoms with the same colours of Fig.~\ref{fig:phase_4}(a) of the main text. This picture is interesting because it clearly shows that, if one focus just on the instantaneous configurations, quantum effects hide the differences between competing geometries of phase IV. 

\subsection*{PIMD simulations analysis}
In this section we add further comments to Tab.~\ref{tab:energies} of the main text, we analyze the average bond length of the molecules and the pair distribution functions for all the simulations carried out in this work. We then report the DOS for the C2/c-24 and Cmca-12 geometries and the vibron modes of mixed structures to analyse the impact of Nuclear Quantum Effects (NQE) with respect to the harmonic theory.
\begin{table}[h!]
\caption{Average bond lengths of the H$_2$ molecule in the different structures. Lengths are given in Bohr units.}
\begin{ruledtabular}
\begin{tabular*}{\textwidth}{{ccc}}\label{tab:lengths}
         {\small \textit{Structure}} & {\small \textit{Pressure (GPa)}} & {\small \textit{Bond length}} \\
\hline
 $C2/c$--$24$ & 200 & 1.3968   \\ 
 $C2/c$--$24$ & 240 & 1.4053 \\ 
 $C2/c$--$24$ & 280 & 1.4145 \\ \\
 $Cmca$--$12$ & 240 & 1.4357  \\ 
 $Cmca$--$12$ & 280 & 1.4443  \\ \\
 $Pbcn$--$48$ & 280 &  1.3684 \\ \\
 $Pc$--$48$ & 280 &  1.3685 \\ \\ 
 $Ibam$--$8$ & 280 &  1.3693 \\ \\
 $Ama2$--$24$ & 280 &  1.3690 \\ \\
\end{tabular*}
\end{ruledtabular}
\end{table}
\begin{figure}
	\includegraphics[scale=0.68]{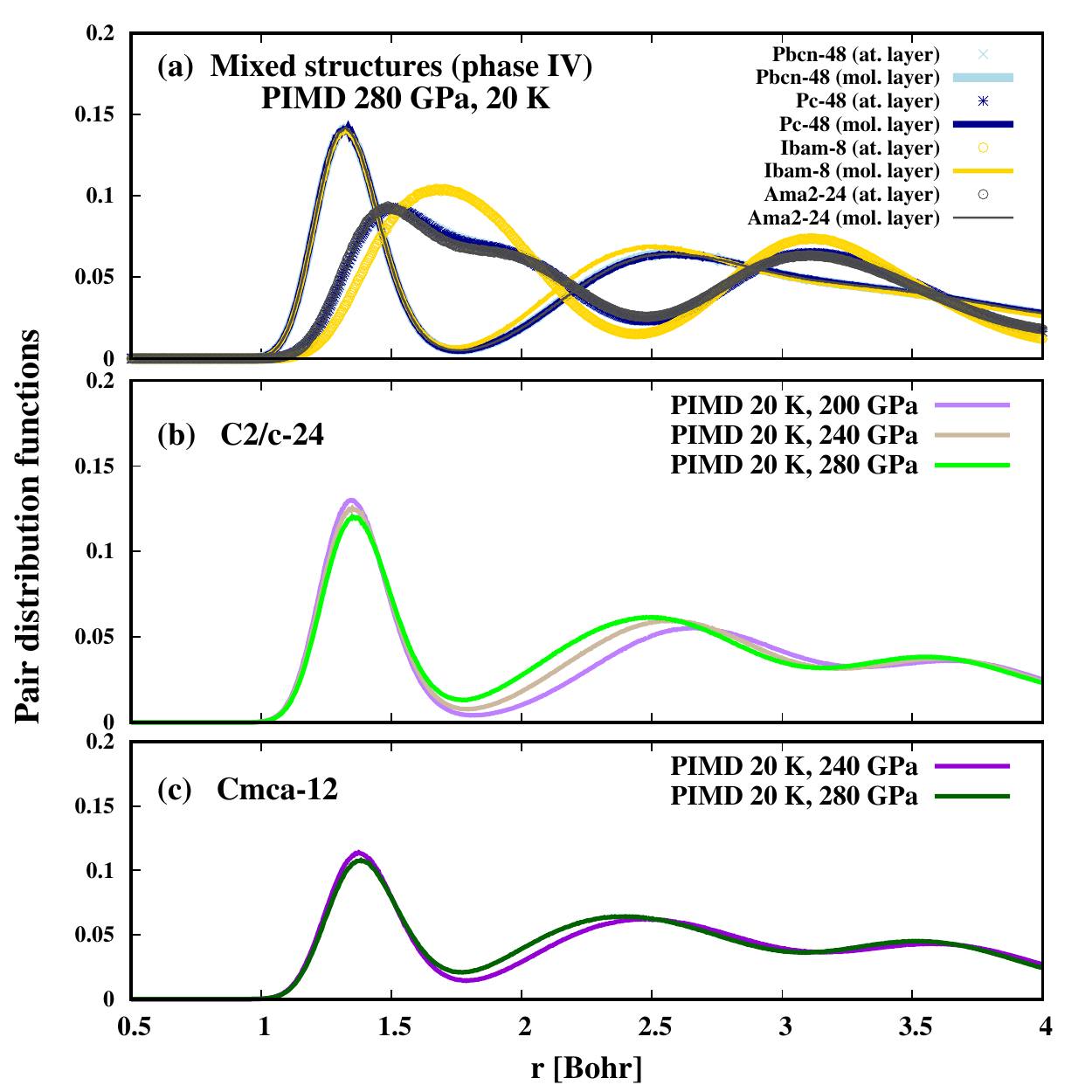}
	\caption{Layered pair distribution functions for the structures studied in this work. (a) Mixed structures (Pbcn-48, Pc-48, Ibam-8 and Ama2-24); (b) C2/c-24; (c) Cmca-12. In (a) we plot both the g(r) belonging to atomic-like planes (using points) and the g(r) belonging to molecular planes (solid lines). A similar analysis but at different temperatures is done in Ref.~\cite{Rillo_2018}.}\label{fig:gr}
\end{figure}
In Tab.~\ref{tab:energies} of the main text we can observe that, for the C2/c-24 and Cmca-12 geometries, the energy expectation value decreases as the pressure increases and that, at 280 GPa and 20 K, the lowest energy structure is the C2/c-24, as already found in different works~\cite{Pickard_2007,McMahon_2012,Drummond_2015}. It is also worth to note that we have attempted to simulate the Cmca-4 geometry~\cite{Drummond_2015} at the same thermodynamic conditions, but after a few ns we observe a phase transition into a mixed layered structure, very close to the Pbcn-48. Therefore, we 
evince that the Cmca-4 structure is unstable at 280 GPa and 20 K using the BLYP functional. The same finding is reported in Ref.~\cite{Rillo_2018} at 200 K. 
In Tab.~\ref{tab:lengths} we report the average bond lengths of H$_2$ molecules within the different geometries (Fig.~\ref{fig:geometries}). For the mixed structures, i.e. Pbcn-48, Pc-48, Ibam-8 and Ama2-24. This analysis is performed by taking into account only layers of strongly bonded molecules. We observe that the bond length increases with pressure in the C2/c-24 and Cmca-12 samples. Molecular bond lengths of the mixed structures are very close to each other and significantly smaller than the ones in purely molecular structures (C2/c-24 and Cmca-12). 
This can be rationalized also by looking at the pair distribution function g(r) in Fig.~\ref{fig:gr}. In Fig.~\ref{fig:gr}(b) and \ref{fig:gr}(c) we observe a small shift of the first peak to the left by increasing pressure for the C2/c-24 and Cmca-12 geometries respectively. In Fig.~\ref{fig:gr}(a) we instead observe an almost exact overlap for the first peak, denoting that the molecular layers of the four different mixed structures 
behave similarly. In the latter case instead we denote a difference in the atomic layer (points in Fig.~\ref{fig:gr}(a)). Indeed, in that case while the Ibam-8 g(r) has a single peak, meaning that during the simulation the hexagonal symmetry of the plane is conserved, for the Pbcn-48, Pc-48 and Ama2-24 we observe a double peak reflecting the presence of different distances between atoms in the atomic-layer. Furthermore, the curves corresponding to the atomic planes Pbcn-48 and Pc-48 
overlap,
making the two structures very hard to distinguish during the PIMD simulation. 
\begin{figure}
	\includegraphics[scale=0.75]{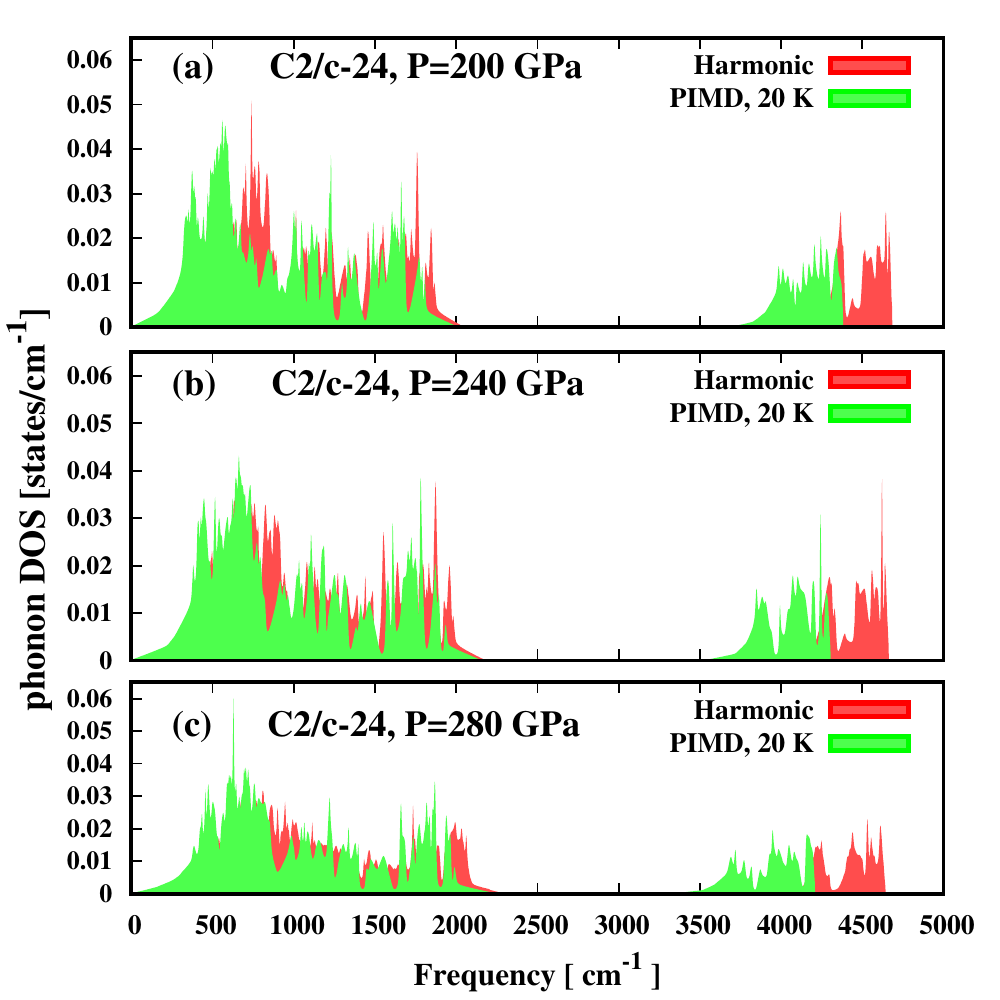}
	\caption{Theoretical phonon DOS of the C2/c-24 geometry for three different pressures: (a) 200 GPa; (b) 240 GPa; (c) 280 GPa. In red we report the harmonic DOS while in green the anharmonic DOS from PIMD simulations. All the DOS are obtained by interpolating the four q-points sampled by the supercell choice and smoothed with a normalized Lorentzian function having a linewidth equal to 10 cm$^{-1}$.}\label{fig:phonon_dos_c2c}
\end{figure}
\begin{figure}
	\includegraphics[scale=0.73]{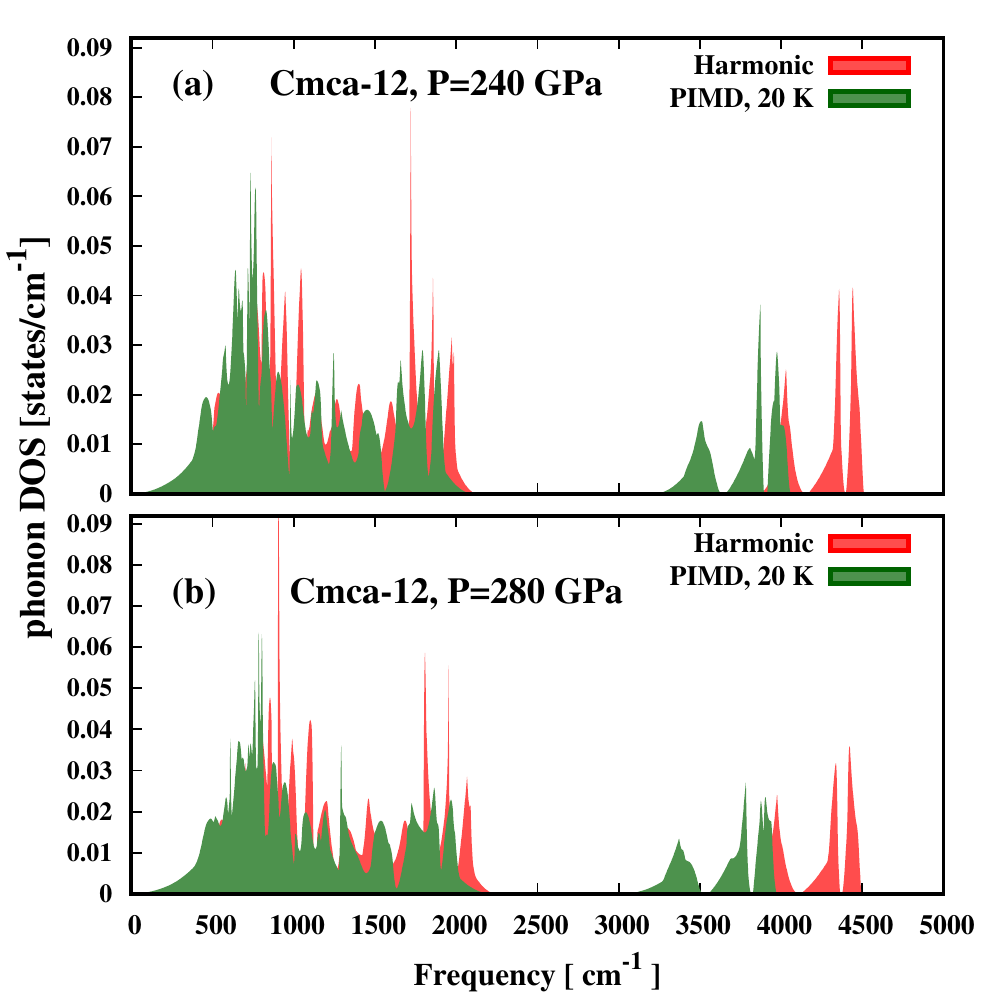}
	\caption{Theoretical phonon DOS of the Cmca-12 geometry for (a) 240 GPa and (b) 280 GPa. As in Fig.~\ref{fig:phonon_dos_c2c} we report the harmonic DOS in red. The anharmonic DOS are given in dark-green.}\label{fig:phonon_dos_cmca}
\end{figure}
\begin{figure*}
	\includegraphics[scale=0.75]{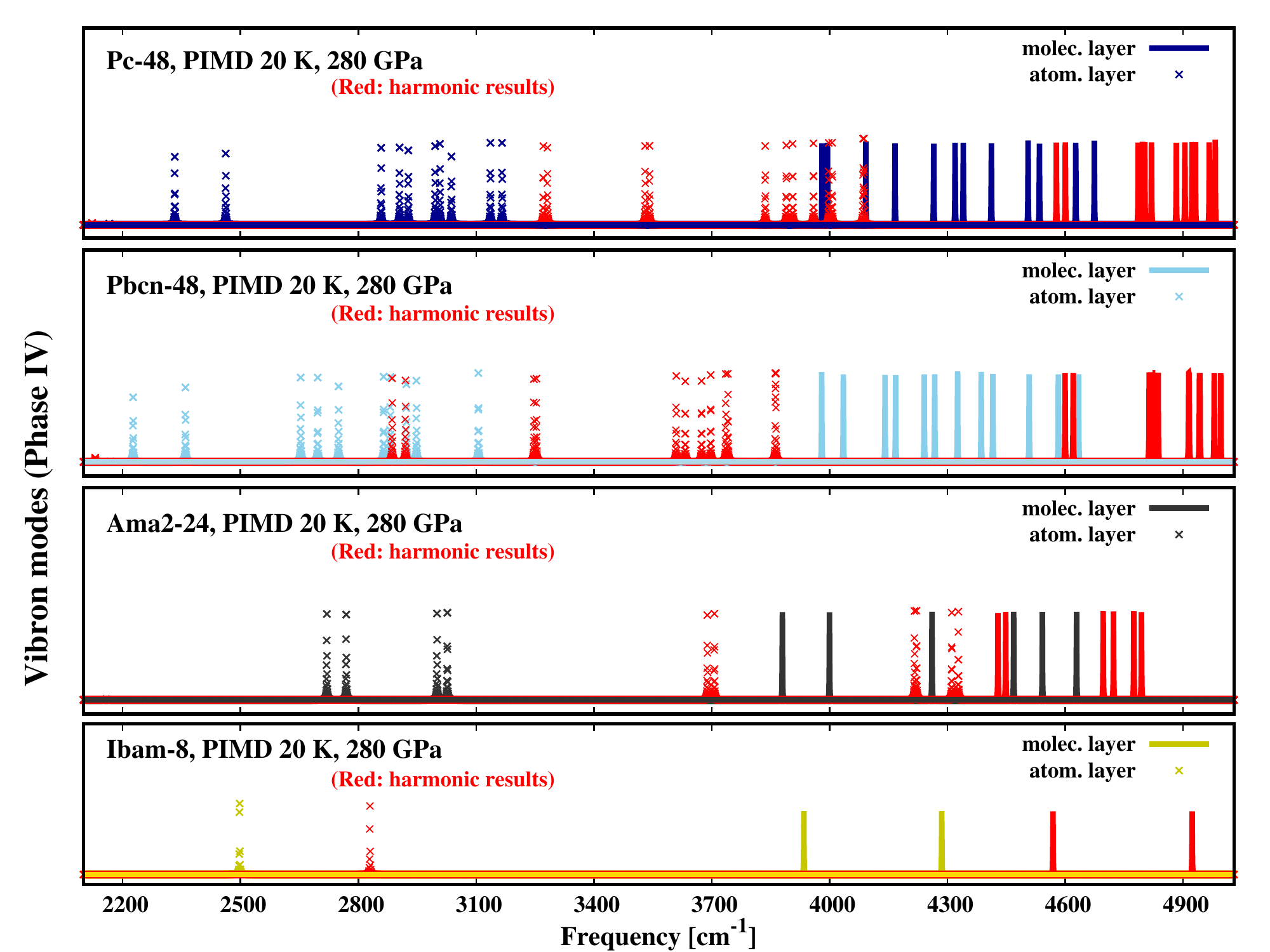}
	\caption{Anharmonic theoretical vibron modes vs harmonic vibron modes (red) for the four different geometries: (a) Pc-48; (b) Pbcn-48; (c) Ama2-24; (d) Ibam-8. Continuous lines represent modes projected over the strongly bonded molecular layers, while points projections over atomic-like layers.}\label{fig:vibrons_phase_4_harm}
\end{figure*}

The harmonic (red) and anharmonic DOS for the pure molecular structures (C2/c-24 and Cmca-12) are reported in Figs.~\ref{fig:phonon_dos_c2c} and \ref{fig:phonon_dos_cmca}. We observe in both cases that, while the lattice modes (modes at frequencies below 2500 cm$^{-1}$) are not heavily affected by NQE, vibron modes are strongly renormalized with respect to the harmonic case. A similar finding was reported in~\cite{Monacelli_2021} for the C2c/-24. 

Finally, a remarkable softening of $\Gamma$-point PIMD phonons with respect to the harmonic case is observed for the mixed-layers structures describing phase IV in Fig.~\ref{fig:vibrons_phase_4_harm}.

\end{document}